# Novel Reversible Multiplier Architecture Using Reversible TSG Gate

Himanshu Thapliyal and M.B Srinivas
*Center for VLSI and Embedded System Technologies*
*International Institute of Information Technology, Hyderabad-500019, India*
E-mail: (thapliyalhimanshu@yahoo.com, srinivas@iiit.net)

**Abstract**

*In the recent years, reversible logic has emerged as a promising technology having its applications in low power CMOS, quantum computing, nanotechnology, and optical computing. The classical set of gates such as AND, OR, and EXOR are not reversible. Recently a 4 * 4 reversible gate called "TSG" is proposed. The most significant aspect of the proposed gate is that it can work singly as a reversible full adder, that is reversible full adder can now be implemented with a single gate only. This paper proposes a NXN reversible multiplier using TSG gate. It is based on two concepts. The partial products can be generated in parallel with a delay of d using Fredkin gates and thereafter the addition can be reduced to $\log_2 N$ steps by using reversible parallel adder designed from TSG gates. Similar multiplier architecture in conventional arithmetic (using conventional logic) has been reported in existing literature, but the proposed one in this paper is totally based on reversible logic and reversible cells as its building block. A 4x4 architecture of the proposed reversible multiplier is also designed. It is demonstrated that the proposed multiplier architecture using the TSG gate is much better and optimized, compared to its existing counterparts in literature; in terms of number of reversible gates and garbage outputs. Thus, this paper provides the initial threshold to building of more complex system which can execute more complicated operations using reversible logic.*

## 1. Introduction

This section provides an effective background of reversible logic with its definition, the motivation behind it, and some key features of the proposed work.

### 1.1. Definitions

Researchers like Landauer have shown that for irreversible logic computations, each bit of information lost generates $kT\ln 2$ joules of heat energy, where k is Boltzmann's constant and T the absolute temperature at which computation is performed [1]. Bennett showed that $kT\ln 2$ energy dissipation would not occur, if a computation is carried out in a reversible way [2], since the amount of energy dissipated in a system bears a direct relationship to the number of bits erased during computation. Furthermore, voltage-coded logic signals have energy of $Esig = \frac{1}{2}CV^2$, and this energy gets dissipated whenever switching occurs in conventional (irreversible) logic implemented in modern CMOS technology. It has been shown that reversible logic helps in saving this energy using charge recovery process [13]. Reversible circuits are those circuits that do not lose information. Reversible computation in a system can be performed only when the system comprises of reversible gates. These circuits can generate unique output vector from each input vector, and vice versa, that is, there is a one-to-one mapping between input and output vectors. Thus, an NXN reversible gate can be represented as

$Iv=(I1,I2,I3,I4,……………………IN)$
$Ov=(O1,O2,O3,………………….ON)$.

Where Iv and Ov represent the input and output vectors respectively. Classical logic gates are irreversible since input vector states cannot be uniquely reconstructed from the output vector states. There are a number of existing reversible gates such as Fredkin gate[3,4,5], Toffoli Gate (TG) [3, 4] and New Gate (NG) [6].

### 1.2. Motivation behind Reversible Logic

The reversible logic operations do not erase (lose) information and dissipate very less heat. Thus, reversible logic is likely to be in demand in high speed power aware circuits. Reversible circuits are of high interest in low-power CMOS design, optical computing, quantum computing and nanotechnology. The most prominent application of reversible logic lies in quantum computers. A quantum computer can be viewed as a quantum network (or a family of quantum networks) composed of quantum logic gates; each gate performs an elementary unitary operation on one, two or more two–state quantum systems called qubits. Each qubit represents an elementary unit of information corresponding to the classical bit values 0 and 1. Any unitary operation is reversible, hence quantum networks effecting elementary arithmetic operations such as addition, multiplication and exponentiation cannot be directly deduced from their classical Boolean counterparts (classical logic gates such as AND or OR are clearly irreversible).Thus, Quantum

Arithmetic must be built from reversible logic components [10].

### 1.3. Proposed Contribution

In this paper, the focus is on the application of new reversible 4*4 TSG gate [12] and its implementation for designing novel reversible multiplier. A NXN reversible multiplier is also proposed in this paper. A similar multiplier in conventional arithmetic (using conventional logic) has been reported in [14]. It is based on two concepts. The partial products can be generated in parallel with a delay of d using Fredkin gates and thereafter the addition can be reduced to $\log_2 N$ steps by using reversible parallel adder designed from TSG gates. A 4x4 architecture of the proposed reversible multiplier is also designed. It has been proved that the proposed multiplier architecture using the proposed TSG gate is better than the existing ones in literature, in terms of number of reversible gates and garbage outputs. The reversible circuits designed and proposed in this paper form the basis of the ALU of a primitive quantum CPU.

## 2. Proposed 4* 4 Reversible Gate

The authors recently proposed a 4*4 one through reversible gate called TS gate (TSG) [12] which is shown in Figure 1. It can be verified that the input pattern corresponding to a particular output pattern can be uniquely determined. The proposed TSG gate is capable of implementing all Boolean functions and can also work singly as a reversible Full adder. Figure 2 shows the implementation of the proposed gate as a reversible Full adder.

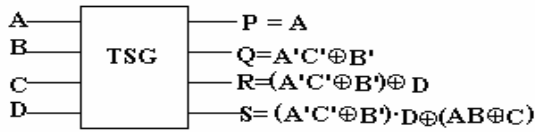

**Figure 1. Proposed TSG Gate**

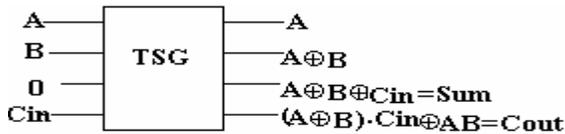

**Figure 2. TSG gate Implemented as a Full Adder**

A number of reversible full adders were proposed in [6,7,8,9]. The reversible full adder circuit in [6] requires three reversible gates (two 3*3 new gates and one 2*2 Feynman gate) and produces three garbage outputs (garbage output refers to the output that is not used for further computations. In other words, it is not used as a primary output or as an input to other gate). The reversible full adder circuit in [7,8] requires three reversible gates (one 3*3 new gate, one 3*3 Toffoli gate and one 2*2 Feynman gate) and produces two garbage outputs. The design in [9] requires five reversible Fredkin gates and produces five garbage outputs. The proposed full adder using TSG in Figure 2 requires only one reversible gate (one TSG gate) and produces only two garbage outputs. Hence, the full-adder design in Figure 2 using TSG gate is better than the previous full-adder designs of [6,7,8,9]. A comparison of various full adders is shown in Table 1.

**Table 1. Comparison of Various Reversible Full Adder Circuits**

|  | Number of Gates | Number of Garbage Outputs | Unit Delay |
|---|---|---|---|
| Proposed Circuit | 1 | 2 | 1 |
| Existing Circuit[6] | 3 | 3 | 3 |
| Existing Circuit [7,8] | 3 | 2 | 3 |
| Existing Circuit[9] | 5 | 5 | 5 |

## 3. Novel Reversible Multiplier Architecture

The proposed reversible NXN bit parallel multiplier architecture is an improvement over reversible array multiplier [11]. Similar multiplier architecture in conventional arithmetic (logic) has been reported in [14], but the proposed one in this paper is totally based on reversible logic and reversible cells as its building block.

It is based on two concepts. The partial products can be generated in parallel with a delay of d using Fredkin gates and thereafter the addition can be reduced to $\log_2 N$ steps by using reversible parallel adder designed from TSG gates. Each two adjacent partial products will be added together with an N-bit reversible parallel adder. A number of interesting and optimized parallel adders are proposed in [12]. The addition of adjacent partial products will generate the first level of computation with N/2 partial sums. These partial sums are added again in the aforesaid fashion to create a second level of computation with N/4 partial sums. The final product will be obtained at the $\log_2 N$ level. The working of the multiplier can be deeply understand by considering a binary tree having N leaf nodes (equivalent to N partial sums) which are merged to form their N/2 parents (equivalent to N/2 Partial Sums). These N/2 parents are again added in the aforesaid fashion and finally this process will be successively repeated to the get at the root of the tree(final product). Thus, the required number of levels to compute the multiplication result will be $\log_2 N$.

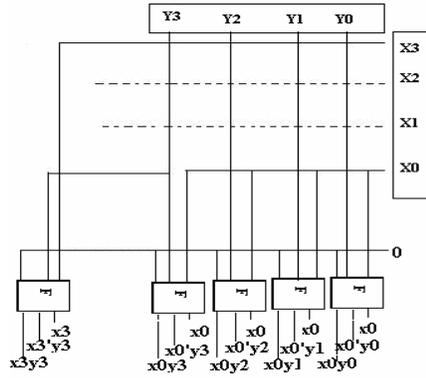

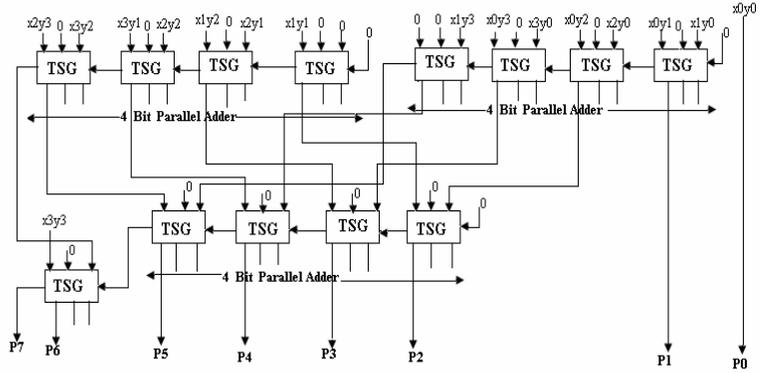

**Figure 3. Parallel generation of Partial Products using Fredkin Gates**

**Figure 5. Proposed 4 x 4 Novel Reversible Multiplier**

Since this architecture requires (N-1) of N-bit reversible adders, it needs a total of N *(N-1) reversible full adder cells. So, the worst propagation delay of the proposed multiplier architecture can be computed as: - d+N*d'[$\log_2 N$] where d and d' are the propagation delays of a Fredkin gate and reversible TSG gate (adder) respectively. By changing the type of adder such as reversible CLA (Carry Look Ahead Adder) to reversible CPA (Carry Propagate Adder) will make a substantial change in the propagation delay.

The proposed NXN reversible multiplier is designed for 4x4 bit. In the 4x4 multiplier, the partial products are generated in parallel using Fredkin gates as shown in Figure 3. Thus, we have 4 partial products generated as shown in Figure 4. Each 2 partial products are added using 4-bit reversible parallel adder creating the first level of computation which has 2 partial sums. These two partial sums are fed to the second level of 4-bit reversible parallel adder, resulting in the formation of the final product. The proposed reversible multiplier efficiency significantly depends on the type of reversible parallel adders used in addition operation. The proposed reversible multiplier is shown in Figure 5 for 4x4 bit. The multiplier uses the proposed TSG gates as reversible full adder units.

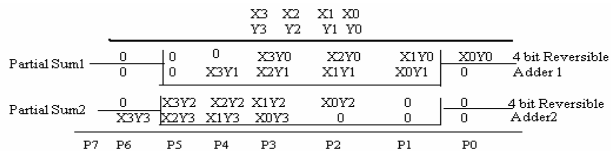

**Figure 4. Methodology of 4x4 reversible Multiplier**

### 3.1. Evaluation of the Proposed Reversible Multiplier

The efficiency of the proposed reversible multiplier greatly depends on the choice of the reversible parallel adder. The efficient parallel adders proposed in [12] will significantly improve the multiplier efficiency. The proposed architecture can further be optimized through a control circuitry. As we have decomposed all the operations into levels, we can significantly reduce the power consumption by employing a control circuitry which will switch off those levels which have done their computations. Therefore, switching off the levels as the computations proceed in the multiplier will lead to a great reduction in power consumption. Furthermore, the multiplier can be optimized for power by having leading zeroes count for both the multiplicand and the multiplier, thus reducing the power consumption by running less number of adders and switching off those adders which are not in use. The multiplier is most optimized compared to its existing reversible counterpart in literature [11]. The proposed 4x4 bit multiplier is designed with bare minimum of 29 reversible gates while its existing counterpart in [11] has 40 reversible gates. The results can be generalized for NXN bits. The numbers of garbage outputs are nearly same for both the multiplier. Table 2 shows the comparison efficiency of the reversible multipliers.

### 4. Conclusions

The focus of this paper is the application of the recently proposed reversible 4*4 TSG gate. A NXN reversible multiplier is also proposed in this paper. It is proved that the proposed multiplier architecture using the proposed TSG gate is better than the existing counterpart in literature in terms of reversible gates and garbage outputs. All the proposed architectures are analyzed in terms of technology independent implementations. The technology independent analysis is necessary since quantum or optical logic implementations are not available. There are a number of significant applications of reversible logics such as low power CMOS, quantum computing, nano-technology, and optical computing and the proposed TSG gate and efficient multiplier architecture are one of the contributions to reversible logic. The proposed circuit can be used for designing large reversible systems. In a nutshell, the advent of reversible logic has contributed significantly in reducing the power consumption. Thus, the paper provides the initial threshold to build more complex systems which can which can execute more complicated operations. The reversible circuits designed and proposed here form the basis of the ALU of a primitive quantum CPU.

**Table 2. Comparison Efficiency of different 4x4 reversible Multipliers**

|  | Proposed Reversible 4x4 Multiplier | Existing 4x4 Counter Parts[11] |
|---|---|---|
| Number of Gates Used | 29 | 40 |
| Switching of the levels (Power Saving) | Yes | No |
| Efficiency depends on reversible Parallel Adder | Yes | No |
| Overall Speed | Fast As Parallel Adder Can be Used | Slow as no parallel Adders |